\documentclass[12pt]{article}
\usepackage{float}
\usepackage{epsfig}

\newcommand{\bfr}{{\bf r}}

\newcommand{\zb}{\bar{z}}
\newcommand{\zt}{\tilde{z}}

\newcommand{\Gt}{\tilde{G}}

\newcommand{\cS}{{\cal S}}
\newcommand{\cG}{{\cal G}}
\newcommand{\ft}{{\tilde f}}
\newcommand{\be}{\begin{equation}}
\newcommand{\ee}{\end{equation}}
\newcommand{\bea}{\begin{eqnarray}}
\newcommand{\eea}{\end{eqnarray}}

\newcommand{\D}{\partial}
\newcommand{\rme}{\mathrm{e}}
\newcommand{\rmi}{\mathrm{i}}
\newcommand{\rmt}{\mathrm{t}}
\newcommand{\rmc}{\mathrm{c}}

\newcommand{\ba}{\begin{array}}   %  
\newcommand{\ea}{\end{array}}     %  
\begin{document}
\baselineskip=17pt

\gdef\journal#1, #2, #3, 1#4#5#6{{#1~}{\bf #2}, #3 (1#4#5#6)}
\gdef\ibid#1, #2, 1#3#4#5{{\bf #1} (1#3#4#5) #2}

\begin{center}
{\Large \bf Exact results for the Spectra of Bosons and Fermions
with Contact Interaction}\\[0.5cm]

{\large \bf Stefan Mashkevich}\footnote{mash@mashke.org}\\[0.1cm]
Schr\"odinger, 120 West 45th St., New York, NY 10036, USA\\[0.2cm]
{\large \bf Sergey Matveenko}\footnote{matveen@landau.ac.ru}\\[0.1cm]
Landau Institute for Theoretical Physics, Kosygina Str. 2,
119334, Moscow, Russia \\[0.2cm]
{\large \bf St\'ephane Ouvry}\footnote{ouvry@lptms.u-psud.fr}\\[0.1cm]
Laboratoire de Physique Th\'eorique et Mod\`eles 
Statistiques\footnote{{\it Unit\'e de
Recherche de l'Universit\'e Paris 11 associ\'ee au CNRS, UMR 8626}}\\
B\^at. 100, Universit\'e Paris-Sud, 91405 Orsay, France
\\
\end{center}

\vskip 0.5cm
\centerline{\large \bf Abstract}
\vskip 0.2cm

An $N$-body bosonic model with delta-contact interactions projected on the
lowest Landau level is considered. For a given number of particles in a given
angular momentum sector, any energy level can be obtained exactly by means of
diagonalizing a finite matrix: they are roots of algebraic equations.
A complete solution of the three-body problem is presented, some general
properties of the $N$-body spectrum are pointed out, and a number of novel
exact analytic eigenstates are obtained. The FQHE $N$-fermion model with
Laplacian-delta interactions is also considered along the same lines of
analysis. New exact eigenstates are proposed, along with the Slater determinant,
whose eigenvalues are shown to be related to Catalan numbers.

\vskip 1cm 
\noindent
PACS numbers: 
71.10-w, 71.70Di, 05.30.Jp\\
%03.65.-w, 05.30.-d, 11.10.-z, 05.70.Ce\\
%Keywords:
%Anyons, Quantum Hall effect, Equation of state

\section{Introduction}

There has recently been considerable interest in Bose-Einstein condensates,
following their experimental discovery in atomic vapors
\cite{Anderson95,Bradley95,Davis95}. Particular attention has been devoted to
fast-rotating condensates, since such systems were demonstrated experimentally
to form vortices \cite{Mattews99,Madison00}, like in superfluid $^4$He;
systems with ever bigger numbers of vortices were subsequently observed
\cite{Madison01,Abo01,Engels02}. A plausible theoretical model describing the
system  is that of two-dimensional (perpendicular to the axis of
rotation) trapped bosons in a harmonic well with delta-function repulsion.
The fast rotation implies that the particles are in a strong effective magnetic field,
thereby allowing for a lowest Landau level (LLL) analysis.
States with a given number of vortices correspond to
the sector of fixed angular momentum $L$ in
the $N$-body problem.

Apart from its application to Bose-Einstein condensates, the problem is
interesting per se, as an $N$-body quantum-mechanical model which can be
solved exactly, at least in part. This model  has  a clear relationship with another
well-known two-dimensional setup in which the classical interaction force is
also zero at nonzero distance --- the anyon model \cite{anyon}. In the latter,
only a small subset of the spectrum can be found exactly
\cite{Nanyons}, whereas within the lowest Landau level, 
 all eigenstates and eigenvalues are explicitly known
\cite{LLLanyons}. One might expect that the delta-interaction LLL problem at
hand would be solvable in the same way, too. But this does not appear to be
the case. An obvious explanation for this difficulty lies in the very
definition of the model itself: delta interactions are known to be
ill-defined in 2 dimensions, since the delta term is too singular with respect to
the kinetic term. This singular interaction has to be regularized in some way.
In fact, the anyon model itself can be viewed as a
regularized version of the delta-interaction model: the long-distance
quantum Aharonov-Bohm interaction, at the heart of the anyonic intermediate
statistics, smoothes out the perturbative divergences due to the short-range contact
interactions also present in the anyon model (exclusion of  the diagonal of the
configuration space). 

Here, in the model considered, the regularization used is the LLL projection
itself. It can be checked that restricting to the LLL subspace allows for a
finite contact interaction to all orders in perturbation theory.

It was conjectured, based on numerical results, that the ground state for
$L\ge N$ has a simple analytical form \cite{Bertsch99}; it was proved later
\cite{Smith00,Papenbrock01} that the state in question is indeed an eigenstate. A
 hierarchy of $N$-body exact eigenstates for $L=4$ was discovered
\cite{Papenbrock01}. The other results are mainly numerical:  the
collective vs single-particle nature of the excited states was explored, good
single-particle approximations in the $L\ll N$ regime discovered
\cite{Papenbrock01}; very recently, a composite-fermion ansatz was suggested,
and it was shown that the CF wave functions at $L=N$ become exact in the
thermodynamic limit \cite{Korslund06}.

In this paper, we concentrate on exact results.
In Sec.~2, we formulate the problem, perform the projection onto the lowest
Landau level, and show that the interaction can be diagonalized separately in
each  given  angular momentum sector. This means that any level
can be obtained by diagonalizing a finite matrix; no basis truncation is
necessary. In Sec.~3, we discuss some general properties of the spectrum and
review a number of cases where the eigenstates have a simple polynomial form
and the corresponding eigenvalues are rational numbers. In the next
section it is shown, by direct counting, that those simple eigenvalues exhaust the
whole spectrum of the three-body problem, which is thus completely solved.
Section 5 deals with the $N$-body problem. Given the nature of the wave
functions involved, it is convenient to consider states with a given value of
$L$ and varying $N$. At $L=2,3$, all the $N$-body eigenstates fall into the category of the
simple states of Sec.~3. For each of $L=4$ and $L=5$, apart from
the simple ones, there is one extra $N$-body eigenstate:  for $L=4$, 
it is the one found in
Ref.~\cite{Papenbrock01}; for $L=5$, it is new. Starting with $L=6$,
most eigenvalues are irrational, but all of them are roots of algebraic
equations. An algorithm is built that lets one find them all. 

Finally, in Sec.~6, a fermionic model introduced so that the Laughlin FQHE
wavefunctions are its ground states is considered, again in the LLL. Via a
mapping onto an equivalent bosonic model, it can be analyzed following the same
approach as in the bosonic case. A~few eigenstates are given as illustrations
of the procedure. The $N$-body Slater determinants, with any $N$, have eigenvalues
which are directly related to Catalan numbers.

\section{Formulation and method of solution}
\label{sec:formulation}

Consider $N$ bosons in two dimensions in a harmonic well
with pairwise contact interaction:
$\hat V= V \sum_{i<j} \delta(\bfr_i-\bfr_j)$.
In complex coordinates ($z=x+\rmi y$), the Hamiltonian is
\be
H = -2\sum_{i=1}^N \D_i \bar\D_i + \frac{\omega^2}{2} \sum_{i=1}^N z_i\zb_i
+ V \sum_{i<j=1}^N \delta^2(z_i - z_j) \;.
\label{eq:H}
\ee

{\bf Projection on the lowest Landau level.} As said above, the delta
interaction is too singular in two dimensions and as such (\ref{eq:H}) is not
properly defined.  One way to regularize the Hamiltonian
is by projecting it on the ``lowest Landau level'' (LLL) subspace.
This means that upon
extracting the long-distance harmonic damping exponential factor
one restricts to the set of $N$-body wave functions 
\be \psi(\bfr_1,
\bfr_2, \ldots , \bfr_N)=\exp\left(-\frac{\omega}{2} \sum_{i=1}^N
z_i\zb_i\right)f(z_1,z_2,...,z_N) \;, \ee 
where $f(z_1,z_2,...,z_N)$ is analytic.

Indeed, if a magnetic field were added, the single-particle eigenstates
corresponding to the LLL (Landau level number $n=0$, angular momentum $l\ge
0$) would be in the LLL-harmonic eigenstates basis

\be \label{6}
\langle z,\bar z|0,l\rangle = \left({{\omega_\rmt ^{l+1}\over \pi l!}}\right)^{{1\over 2}}z^{l}
\rme^{-{1\over 2}\omega_\rmt z\bar z} \;,
\ee
where $\omega_\rmt = \sqrt{\omega_\rmc^2 + \omega^2}$, $\omega_\rmc$ being
half the cyclotron frequency. Coming back to the purely harmonic problem,
consider now the projector on the ``LLL'' Hilbert space 
$ P_0=\sum_{l=0}^{\infty}|0,l\rangle \langle 0,l|$ 
\be \label{8} \langle z,\bar z|P_0|z',\bar z'\rangle ={\omega \over \pi}
\rme^{-{\omega\over 2}(z\bar z+z'\bar z'-2z\bar z')}\ee
A generic state belonging to the LLL,
$|\psi\rangle =\sum_{l=0}^{\infty}a_l|0,l\rangle $,  is analytic up to
the  Gaussian factor,
\be\label{10} \langle z|\psi\rangle =f(z)\rme^{-{\omega\over 2}z\bar z}\ee
with
\be \label{11} f(z)=\sum_{l=0}^{\infty}a_l z^l \;.\ee

Projecting  a one-body Hamiltonian  on the LLL amounts to
\be \label{12} \langle z,\bar z|P_0HP_0|\psi\rangle =
\langle z,\bar z|P_0H|\psi\rangle =\int dz'd\bar z'{\omega \over \pi}
\rme^{-{\omega\over 2}(z\bar z+z'\bar z'-2z\bar z')}H(z')f(z')\rme^{-{\omega \over 2}z'\bar z'}\;,
\ee
that is to say, for a one-body potential $V(z,\bar z)$,  to the eigenvalue equation
\be
\label{14}\int dz'd\bar z'{\omega\over \pi}
\rme^{-{\omega }(z'\bar z'-z\bar z')}
\left[ \omega +\omega z' \partial' + V(z',\bar z')\right]f(z')=Ef(z)\;.\ee

In the $N$-body case with $\hat V=V\sum_{i <j} \delta^2 (z_i - z_j)$ one
obtains a product of $N$ integrals of
$f(z_1,... ,z_N)$. Using the Bargmann identity
\be
\label{barg}{\omega\over \pi}\int dz' d\bar z'
\rme^{-{\omega}(z'\bar z'-z\bar z')}h(z')=h(z) \;,
\ee
after integrations, the projected Hamiltonian reads
(for simplicity, $\omega= 1$ and $V/(2\pi) = 1$)
\be
H_{\rm LLL} = \hat L+ \hat V \;,
\label{eq:LH}
\ee
where the angular momentum operator is
\be
\hat L = \sum_{i=1}^N z_i \D_i
\label{eq:Lop}
\ee
and the LLL-projected contact interaction
\be
\hat V = \sum_{i<j=1}^N P_{ij} \;,
\label{eq:Ht}
\ee
$P_{ij}$ being the operator\footnote{For $\hat V=
 V \sum_{i<j<k=1}^N\delta^2(z_i - z_j) \delta^2(z_i - z_k)$,
when the Pfaffian states are known to play a central role, one would have in the LLL
$$
\hat V = \sum_{i<j<k=1}^N P_{ijk} \;,
$$
where $P_{ijk}$ is the operator that replaces $z_i$, $z_j$  and $z_k$
in the wave function $f\left(\ldots, z_i, \ldots, z_j, \ldots, z_k,\ldots\right)$ with the coordinate of the center of mass
of the triplet $(ijk)$, keeping all the other coordinates intact:
$$
P_{ijk} f\left(\ldots, z_i, \ldots, z_j, \ldots, z_k,\ldots\right)
=\nonumber\\ f\left(\ldots, \frac{z_i+z_j+z_k}{3}, \ldots, \frac{z_i+z_j+z_k}{3}, \ldots
, \frac{z_i+z_j+z_k}{3}, \ldots\right) \;.
$$} 
that replaces both $z_i$ and $z_j$
in the wave function with the coordinate of the center of mass
of the pair $(ij)$, keeping all the other coordinates intact:
\be
P_{ij} f\left(\ldots, z_i, \ldots, z_j, \ldots\right)
= f\left(\ldots, \frac{z_i+z_j}{2}, \ldots, \frac{z_i+z_j}{2}, \ldots \right) \;.
\label{eq:Pij}
\ee

A generic bosonic eigenstate of $\hat L$ is a symmetrized 
monomial in the $z_i$'s:
\be
f_{l_1\ldots l_N} = \cS\left[ \prod_{i=1}^N z_i^{l_i} \right] \;,
\label{eq:psil1lN}
\ee
with
\be
L = \sum_{i=1}^N l_i \;,  \quad  l_i \ge l_{i+1} \ge 0 \;.
\label{eq:Lsum}
\ee
Here the symmetrization operator acts as
\be
\cS[f(z_1,z_2, \ldots, z_N)] = \frac{1}{N!} \sum_{\cal P} f(z_{{\cal P}_1},
z_{{\cal P}_2}, \ldots, z_{{\cal P}_N}) \;,
\label{eq:N}
\ee
the sum being over all $N!$ permutations $\cal P$
of $(1, \ldots, N)$.

The multiplicity $G(N, L)$ of a level with given values of $N$ and $L$
is the number of ways that $L$ can be represented as a sum (\ref{eq:Lsum}),
i.e., the number of unordered partitions of $L$ into $N$ nonnegative addends.
[E.g., for $N=3$, $L=6$, the $G(3,6)=7$ partitions are
$(6,0,0)$,
$(5,1,0)$,
$(4,2,0)$,
$(4,1,1)$,
$(3,3,0)$,
$(3,2,1)$,
$(2,2,2)$.]
It is directly connected to a well-studied quantity $p_N(M)$,
the number of unordered partitions of $M$ into $N$ positive addends:
By adding 1 to each addend in each partition contributing to $G(N, L)$,
one sees that $G(N, L) = p_N(L+N)$.
This is a polynomial of degree $(N-1)$ in $L$:
$G(N, L) = L^{N-1}/[N!(N-1)!] + O(L^{N-2})$.
The generating function is
\be
\prod_{k=1}^\infty \frac{1}{1 - x^k z} = 1 + \sum_{M,N=1}^\infty G(N,M-N) x^M z^N \;.
\ee

The problem reduces to the diagonalization of $\hat V$
in the $G(N, L)$-dimensional subspace spanned by $\{ f_{l_1\ldots l_N} \}$,
for any given $N$ and $L$.
A generic state belonging to that subspace is
\be
f(z_1,z_2,....,z_N) = \sum_{\{l_k\}} a_{l_1\ldots l_N}f_{l_1\ldots l_N}
\label{eq:psial}
\ee with $f_{l_1\ldots l_N}$ defined by Eq.~(\ref{eq:psil1lN}), $a_{l_1\ldots
l_N}$ arbitrary coefficients, and the sum being over all $G(N,L)$ possible
sets $\{l_k\}$ defined by Eq.~(\ref{eq:Lsum}). The eigenvalue equation is
\be \hat V f = Ef \; \ee 
[the total energy, according to (\ref{eq:LH}), is $L+E$]. One has
\be \hat V f =
\sum_{i<j=1}^N P_{ij} f = \sum_{\{l_k\}} a_{l_1\ldots l_N} \sum_{i<j=1}^N
P_{ij}f_{l_1\ldots l_N} \;,
\label{eq:Htpsi}
\ee
and upon expanding all the powers of $\frac{z_i + z_j}{2}$ in $P_{ij}f_{l_1\ldots l_N}$,
this becomes a polynomial in the $z_i$'s.
In Eq.~(\ref{eq:psial}), with account for (\ref{eq:psil1lN}),
$a_{l_1\ldots l_N}$ is the coefficient
at $\prod_k z_k^{l_k}$.
Hence, in the polynomial (\ref{eq:Htpsi}), the coefficient
at $\prod_k z_k^{l_k}$ must be equated to $E a_{l_1\ldots l_N}$,
for each $\{l_k\}$; one obtains $G(N,L)$ equations for as many coefficients $a_{l_1\ldots l_N}$.
Every level can thus be found exactly by
diagonalization of a {\em finite} matrix, without the usual errors induced
by basis truncation.

\section{General properties}
\label{sec:simple}

{\bf Center of mass.} Introduce the center-of-mass coordinate
\be
Z = \frac1N \sum_{i=1}^N z_i
\label{eq:Z}
\ee
and the relative coordinates
\be
\zt_i = z_i - Z \;;
\label{eq:zti}
\ee
obviously,
\be
\sum_{i=1}^N \zt_i = 0 \;.
\label{eq:sumzt}
\ee

For any function of the center-of-mass coordinate, $P_{ij}F(Z) = F(Z)$, hence
\be
\hat V F(Z) = \frac{N(N-1)}{2}F(Z) \;.
\ee
Moreover, for any eigenfunction $f(z_1,z_2,...,z_N)$ of the interaction Hamiltonian,
\be
\hat V f(z_1,z_2,...,z_N) = Ef(z_1,z_2,...,z_N) \;,
\ee
also
\be
\hat V F(Z)f(z_1,z_2,...,z_N) = EF(Z)f(z_1,z_2,...,z_N) \;
\ee
--- a center-of-mass excitation does not change the interaction
energy (although it does affect the total energy
because the angular momentum changes).
Thus, for each $(N, L)$ state, there is a ``tower'' of CM excitations above it:
$(N, L+1)$, $(N, L+2)$, etc. It is enough to find the ``pure relative''
states, whose number is
\be
\Gt(N, L) = G(N, L) - G(N, L-1) \;.
\label{eq:Gt}
\ee

{\bf Known exact eigenstates.} In the $N$-body case, denote
\be
{\tilde f}_{l_1\ldots l_N} = \cS \left[ \prod_{i=1}^N \zt_i^{l_i} \right] \;.
\ee
One has
\bea \hat V {\tilde f}_{1 \ldots 1} & = & \sum_{i<j=1}^N \left(\frac{\zt_i +
\zt_j}{2}\right)^2 \prod_{\stackrel{\scriptstyle k = 1}{k \ne i,j}}^N \zt_k =
\frac14 \sum_{i<j=1}^N \frac{(\zt_i + \zt_j)^2}{\zt_i \zt_j} \prod_{k = 1}^N
\zt_k \nonumber \\ & = & \frac14 \sum_{i<j=1}^N \left( 2+\frac{\zt_i}{ \zt_j}
+ \frac{\zt_j}{\zt_i} \right) \prod_{k = 1}^N \zt_k = \frac14 \left[ N(N-1) +
\sum_{i \ne j=1}^N \frac{\zt_i}{\zt_j} \right] \prod_{k = 1}^N \zt_k \nonumber
\\ & = & \frac14 \left[ N(N-1) - N + \sum_{i,j=1}^N \frac{\zt_i}{\zt_j}
\right] \prod_{k = 1}^N \zt_k = \frac{N(N-2)}{4} \, {\tilde f}_{1 \ldots 1}
\label{eq:psi111}
\eea
[at the very last step, Eq.~(\ref{eq:sumzt}) was used].

More generally, for $1 < L \le N$,
\bea \hat V {\tilde f}_{\underbrace{\scriptstyle 1 \ldots 1}_L0 \ldots 0} & = &
\frac{N}{2} \left( N - \frac{L}{2} - 1 \right){\tilde f}_{1 \ldots 10 \ldots 0}
\label{eq:psi110} 
\eea
(see Refs.~\cite{Smith00,Papenbrock01} for two different proofs).
It has been conjectured numerically 
that this is the ground state for any $L\le N$ \cite{Bertsch99}.
Obviously, for $L=N$ it reduces to (\ref{eq:psi111}).
Note also that ${\tilde f}_{10\ldots 0} \equiv 0$ because of Eq.~(\ref{eq:sumzt}).

Further, in the $3$-body case, for any $L > 1$, one has 
\bea
\hat V {\tilde f}_{L00} & = &
\sum_{i<j=1}^3 P_{ij} \frac13 \sum_{k=1}^3 \zt_k^L \nonumber \\
& = & \frac13 \sum_{i<j=1}^3 \left[ 2 \left(\frac{\zt_i + \zt_j}{2}\right)^L - \zt_i^L
- \zt_j^L  \right] + 3{\tilde f}_{L00}   \nonumber \\
% & = & \left[ \frac{(N-1)(N-2)}{2} + \frac{(-1)^L (N -1)}{2^L}\right] {\tilde f}_{L0 \ldots 0}
& = & \left[ 1+ \frac{(-1)^L }{2^{L-1}} \right] {\tilde f}_{L00} \;,
\label{eq:psiL00}
\eea
since  $\zt_i + \zt_j=-\zt_k$ for $i\ne j\ne k$.

In the $4$-body case, for any odd $L$,
\be
\hat V{\tilde f}_{L000} = 3 {\tilde f}_{L000} \;,
\label{eq:fL000}
\ee
since in this case $\zt_i + \zt_j=-\zt_k-\zt_l$ for $i\ne j\ne k\ne l$.   

In the $N$-body case, for $L=2$ and $L=3$,
respectively, $\ft_{20\ldots0}$ and $\ft_{30\ldots0}$
are eigenstates of the form (\ref{eq:psi110})
since, trivially, $\sum_{k=1}^N \zt_k^2=-\sum_{k,l}^N \zt_k\zt_l$
and, less trivially, $\sum_{k=1}^N \zt_k^3=
\sum_{k,l,m}^N \zt_k\zt_l\zt_m$.

{\bf ``Slater excitations''.} Consider the square of
the $N$-particle Slater determinant,
\be
S_N^2 = \prod_{i<j=1}^N (z_i - z_j)^2= \prod_{i<j=1}^N (\zt_i - \zt_j)^2\; .
\ee
Obviously, $P_{ij}S_N^2 f = 0$ for any $f(z_1,\ldots,z_N)$.
Hence, a ``Slater excitation'' of any function is an eigenfunction
of $\hat V$ with zero eigenvalue:
\be
\hat V S_N^{2n}f(z_1,z_2,...,z_N) = 0 \;.
\label{eq:S2n}
\ee
(The presence of a Slater determinant nullifies
the probability for the positions of any two particles
to coincide, hence the delta interaction has no effect.)
An even power of the determinant is required
in order to preserve the symmetry of the function.
This is somewhat reminiscent of the idea of ``composite fermions'' \cite{Korslund06}.

\section{Three-body problem}

For $N=3$, it turns out that the simple eigenstates
described in the previous section exhaust the whole spectrum.

The number of pure relative eigenstates for any $L\ge0$ is
\be
\Gt(3,L) = \left[ \frac{L}{6} \right] + 1 - \delta_{L\:\mathrm{mod}\:6, 1} \;,
\ee
i.e., $\Gt(3,0) = 1$,  $\Gt(3,1) = 0$, $\Gt(3,2) = \ldots = \Gt(3,5) = 1$, $\Gt(3,6) = 2$,
and so on with period 6.
Here is the systematics of these states.
For any $L \ne 1$, there
is a state ${\tilde f}_{L00}$ with the eigenvalue
$1 + (-1)^L/2^{L-1}$, Eq.~(\ref{eq:psiL00}).
These make up all the states for $L \le 5$.
[Recall that ${\tilde f}_{110}$, Eq.~(\ref{eq:psi110}), is the same as
$\ft_{200}$.]
Now, upon any $(3,L)$ state $\ft$ there is a tower of $(3,L+6n)$ Slater
excitations $S_3^{2n}{\tilde f}$, $n=1,2,\ldots$, with zero eigenvalues.
These excitations account for the fact that $\Gt(3,L+6) = \Gt(3,L) + 1$
[at $(3,L+6)$ there is one new state, ${\tilde f}_{L+6,00}$,
plus Slater excitations of all the $(3,L)$ states]
and complete the relative spectrum.

As an illustration, all the pure relative $(3,L)$ eigenstates
for up to $L=17$ are enumerated below:

\begin{center}
\begin{tabular}{|c|l||c|l||c|l|}
\hline
$L$ & $E$ & $L$ & $E$ & $L$ & $E$ \\ \hline
0 & 3 & 6 & $\frac{33}{32}$, 0 & 12 & $\frac{2049}{2048}$, 0, 0 \\ \hline
1 & --- & 7 & $\frac{63}{64}$ & 13 & $\frac{4095}{4096}$, 0 \\ \hline
2 & $\frac32$ & 8 & $\frac{129}{128}$, 0 & 14 & $\frac{8193}{8192}$, 0, 0 \\
\hline
3 & $\frac34$ & 9 & $\frac{255}{256}$, 0 & 15 & $\frac{16383}{16384}$, 0, 0 \\
\hline
4 & $\frac98$ & 10 & $\frac{513}{512}$, 0 & 16 & $\frac{32769}{32768}$, 0, 0 \\
\hline
5 & $\frac{15}{16}$ & 11 & $\frac{1023}{1024}$, 0 & 17 & $\frac{65535}{65536}$, 0, 0
\\ \hline
\end{tabular}
\end{center}

Adding a tower of center-of-mass excitations to each of these,
one gets the complete 3-body (LLL) spectrum.

In the spirit of Eq.~(\ref{eq:S2n}), one could also consider
``generalized Slater determinants'', of the form
\be
T_{2n_1,2n_2,2n_3} =
\cS \left[ (\zt_1 - \zt_2)^{2n_1} (\zt_1 - \zt_3)^{2n_2}
(\zt_2 - \zt_3)^{2n_3} \right] \;;
\ee
obviously, $T_{2n_1,2n_2,2n_3}$ times any symmetric function
is a bosonic state annihilated by $\hat V$.
However, this does not yield new states. E.g.,
\be
T_{422} = 3 S_3^2 \ft_{200} \;; \quad
T_{642} = \frac{243}{22} S_3^2 \ft_{600} - \frac{1}{11} S_3^4 \;.
\ee

\section{$N$-body problem}

As already stated, the diagonalization of the interaction Hamiltonian
can be performed separately in each $(N,L)$ sector.
An improvement can be made by excluding the center-of-mass
excitations {\em a priori}, i.e., diagonalizing in the pure relative basis
--- thereby reducing the dimension of the subspace involved
from $G(N,L)$ to $\Gt(N,L)$.
Since the maximum possible number of nonzero addends in a partition of $L$
is $L$ itself, $G(N,L)$ ceases to grow with $N$ at $N = L$.
Hence, for any $N \ge L$ also $\Gt(N,L) = \Gt(L,L) \equiv \cG(L)$.
Now, by definition (\ref{eq:Gt}), $\cG(L) = G(L,L) - G(L,L-1)$.
In each of the $G(L,L-1)$ unordered partitions
of $L-1$ into $L$ nonnegative addends, at least one addend is equal to zero.
Replacing that zero with one, we see that
$G(L,L-1)$ is also the number of unordered partitions
of $L$ into $L$ nonnegative addends of which at least one is equal to one.
Hence, $\cG(L)$ is the number of unordered partitions
of $L$ into $L$ nonnegative addends of which {\em none} is equal to one.
E.g., $\cG(6) = 4$, which corresponds to the four
partitions (zeroes dropped for brevity): $(6)$,
$(4,2)$, $(3,3)$, $(2,2,2)$.
The generating function is
\be
\prod_{k=1}^\infty \frac{1}{1 - x^k} = 1 + \sum_{L=1}^\infty \cG(L) x^L \;.
\ee
\begin{center}
\begin{tabular}{|c|c|c|c|c|c|c|c|c|c|c|c|c|}
\hline
$L$ & 1 & 2 & 3 & 4 & 5 & 6 & 7 & 8 & 9 & 10 & 11 & 12 \\
\hline
$\cG(L)$ & 0 & 1 & 1 & 2 & 2 & 4 & 4 & 7 & 8 & 12 & 14 & 21 \\
\hline
\end{tabular}
\end{center}

Thus, all the pure relative eigenstates for any $N$ at a given $L$
can be found by diagonalizing no more than a $\cG(L) \times \cG(L)$ matrix.
The basis is formed by the set of states
$\{ \ft_{l_1\ldots l_N} :\, \sum_{i=1}^N l_i = L,\; l_i \ge l_{i+1} \ge 0,\; l_i \ne 1\}$.
Indeed, any state with any number of $l_i$'s equal to 1
is a linear combination of the basis states:
Since $\zt_i = -\sum_{j\ne i}^N \zt_j$,
\bea
\ft_{l_1 \ldots l_{i-1},1,l_{i+1} \ldots l_N} & = &
-\left(
\ft_{l_1+1 \ldots l_{i-1},0,l_{i+1} \ldots l_N} + \ldots 
+ \ft_{l_1 \ldots l_{i-1}+1,0,l_{i+1} \ldots l_N} \right. \nonumber \\
&& \left. {} + \ft_{l_1 \ldots l_{i-1},0,l_{i+1}+1 \ldots l_N} + \ldots
+ \ft_{l_1 \ldots l_{i-1},0,l_{i+1} \ldots l_N+1} \right) \;,
\eea
and if some $l_k=0$, then the addend involving $l_k+1$ is equal
to the LHS, due to symmetry.
A single $l_i=1$ is thus eliminated,
and by repeating, one eliminates them all.

One pure relative eigenstate for any $L$ and $N \ge L$ is already known, Eq.~(\ref{eq:psi110}).
For $L=2$ and $L=3$, as evidenced by the table above, there are no more.
For $L=4$, apart from the state (\ref{eq:psi110}), which in this case is
\bea
\ft_{N,4}^{(1)} = \sum_{i \ne j \ne k \ne l}^N \zt_i\zt_j\zt_k\zt_l \;,
\eea
with the eigenvalue
\bea
E_{N,4}^{(1)} = \frac{N(N-3)}{2} \;,
\eea
there is the state 
\be 
\ft_{N,4}^{(2)} = \sum_{i\ne j}^N({\tilde z}_i-{\tilde z}_j)^4 \;,
\ee
whose eigenvalue is
\be
E_{N,4}^{(2)} = {N(N-1)\over 2}-{7\over 8}N+{3\over 4} \;,
\label{eq:EN42}
\ee
as found in Ref.~\cite{Papenbrock01}.

For $L=5$, the eigenstate
\bea
\ft_{N,5}^{(1)} = \sum_{i \ne j \ne k \ne l\ne m}^N \zt_i\zt_j\zt_k\zt_l\zt_m \;,
\eea
\bea
E_{N,5}^{(1)} = \frac{N(2N-7)}{4} \;
\eea
is complemented by a new eigenstate
\be
\ft_{N,5}^{(2)} = \sum_{i\ne j\ne k}^N({{\tilde z}_i+{\tilde z}_j\over 2}-{\tilde z}_k)^5
\ee
with
\be
E_{N,5}^{(2)} = {N(N-1)\over 2}-{15\over 16}N+{3\over 4} \;.
\label{eq:EN52}
\ee

All the $N$-body eigenstates with  $L\le5$ have thus been found exactly.
Equations (\ref{eq:EN42}) and (\ref{eq:EN52}) seem to suggest a systematic
pattern --- which, however, does not exist.

For $L=6$, the basis states are
\bea 
h_1 & = & \sum_i^N {\tilde z}_i^6 \;, \\
h_2 & = & \sum_{i \ne j}^N {\tilde z}_i^4 {\tilde z}_j^2 \;, \\
h_3 & = & \sum_{i< j}^N{\tilde z}_i^3 {\tilde z}_j^3 \;, \\
h_4 & = & \sum_{i \ne j \neq k}^N {\tilde z}_i^2 {\tilde z}_j^2 {\tilde z}_k^2 \;;
\eea
defining a matrix $V_{ij}$  such that
\be \hat{V} h_i = \sum_j V_{ij}h_j \;,\ee
one obtains
\be
||V_{ij}|| = \left( \begin{array}{cccc}
\frac{16N^2 - 47N + 25}{32} & \frac{15}{32} & \frac{5}{8} & 0 \\
\frac{N+9}{32} & \frac{16N^2 - 60N + 39}{32} & -\frac{3}{8} & \frac{3}{8} \\
\frac{N-7}{64} & -\frac{33}{64} & \frac{8N^2 - 32N + 13}{16} & 0 \\
0 & \frac{3N+6}{8} & 3 & \frac{4N^2 - 16N + 9}{8}
\end{array} \right) \;.
\ee
One eigenvalue is
\be
E_{N,6}^{(1)} = \frac{N(N-4)}{2} \;,
\label{eq:EN61}
\ee
which is nothing but Eq.~(\ref{eq:psi110}) with $L=6$;
the other three eigenvalues are the roots of the equation
\bea
&& 512E^3 + (-768 N^2 + 2736 N - 2016)E^2 \nonumber \\
&& \quad {} + (384 N^4 -2736 N^3 + 6850 N^2 - 7116 N + 2664)E \nonumber \\
&& \quad {} + (-64 N^6 + 684 N^5 - 2921 N^4 + 6378 N^3 - 7527 N^2 + 4554 N - 1080) = 0 \;.
\label{eq:Eirrat}
\eea

For $N<L$, there are fewer than $\cG(L)$ states,
but the general scheme of calculation remains the same.
At $L=6$, the rational eigenvalue
(\ref{eq:EN61}) does not exist for $N<6$,
but the irrational ones do for all $N \ge 4$.
The table below lists all the pure relative eigenvalues
for $4 \le N \le 6$ and $L \le 6$.

\begin{center}
\begin{tabular}{|l|l|l|l|}
\hline
$L$ & $N=4$ & $N=5$ & $N=6$\\ \hline
$2$ & 4 & $\frac{15}{2}$ & 12 \\ \hline
$3$ & 3 & $\frac{25}{4}$ & $\frac{21}{2}$ \\ \hline
$4$ & 2, $\frac{13}{4}$ & 5, $\frac{51}{8}$ & 9, $\frac{21}{2}$\\ \hline
$5$ & 3 & $\frac{15}{4}$, $\frac{97}{16}$ & $\frac{15}{2}$, $\frac{81}{8}$ \\ \hline
$6$ & \small 1.38187, 2.14669, 3.03395 &
\small 3.81328, 4.85852, 6.04695 & \small 6, 7.26393, 8.54438, 10.06670 \\ \hline
\end{tabular}
\end{center}

There appear a few more rational eigenvalues beyond $L=6$.
In the 4-body problem, we already know about the eigenvalue 3
for any odd $L$, Eq.~(\ref{eq:fL000});
besides, one of the (4,7) eigenvalues is $\frac{11}{8}$,
while at (4,12), there duly appears the Slater determinant squared,
$$
S_4^2 = (\zt_1 - \zt_2)^2 (\zt_1 - \zt_3)^2 (\zt_1 - \zt_4)^2
(\zt_2 - \zt_3)^2 (\zt_2 - \zt_4)^2 (\zt_3 - \zt_4)^2 \;,
$$
with eigenvalue 0, Eq.~(\ref{eq:S2n}).
Most of the values, however, appear to be irrational.

\section{A model of interacting fermions}

Consider now a Fermi gas with the interaction
\be \hat V=V\sum_{i < j} \Delta_i \delta (\bf{r}_i - \bf{r}_j ) \;,\ee
which is well-known to have a trivial ground state --- the Laughlin FQHE wave functions
(with the vanishing eigenvalue) \cite{Haldane}.

Let us again consider the model projected on the ``LLL'': 
it transforms into (again, $\omega=1$  and $V/(2\pi)=1$)  
\be\label{toto}  \hat V=\sum_{i < j}(z_i - z_j ) \left[\left.\frac{\partial}{\partial z_i}
f (z_1, \ldots, z_N )\right\vert_{z_i, z_j \to (z_i + z_j )/2 }\right] \;.
\label{ferm}
\ee
Necessarily, one has  $f(z_1, \ldots, z_N) = S_N h(z_1, \ldots, z_N)$, where
$h(z_1, \ldots, z_N)$ is a regular $N$-body bosonic wave function. 
In turn, (\ref{ferm}) acting on
$h(z_1, \ldots, z_N)$ takes the form
\be\label{totobis} \sum_{i < j} {\left.\left[\frac{S_N}{z_i - z_j} h \right] \right
\vert_{z_i, z_j \to
(z_i + z_j )/2 }\over \frac{S_N}{z_i - z_j}} \;. \ee
Therefore, one is back to a situation quite reminiscent of the bosonic problem
addressed above, and the same machinery applies, but with the new
Fermi LLL-interaction term given by Eq.~(\ref{totobis}).

One almost obvious $N$-body eigenstate is
$S_N = \prod_{i < j}^N (z_i - z_j )$ [i.e., $h(z_1,\ldots,z_N)=1$].
Indeed, the interaction term (\ref{toto}) maps a fermionic state
with a given angular momentum on a fermionic state with the same angular
momentum without introducing any singularity.
It follows that, in the lowest possible angular momentum sector $L=N(N-1)/2$,
the single state $S_N$ is necessarily mapped onto itself.
This is expressed in the identity
\be \sum_{i < j} {\left.\left[\frac{S_N}{z_i - z_j} \right] \right\vert_{z_i, z_j \to
(z_i + z_j )/2 }\over\frac{S_N}{z_i - z_j}}  = E_N \;,\ee
or
\be\label{paul}\sum_{i < j} \prod_{k \neq i, j}
\frac{[(z_i - z_k) + (z_j - z_k)]^2}{(z_i - z_k)(z_j - z_k)}
= D_{N-1} \;,\ee
where
\be
E_N = \frac{D_{N-1}}{2^{2(N-2)}} \;
\ee
and
\be
D_N = \frac{N+1}{2} \left[ 4^N - {2N+1 \choose N} \right]
\ee
is related to the Catalan integer sequence $C_j$:
\be
D_N = \sum_{j=0}^{N-1} C_j (N-j) 4^{N-j-1} \;, \qquad C_j = \frac{1}{j+1} {2j \choose j} \;.
\ee

The identity can be shown to hold in general by relating $D_{N-1}$ 
to the double contour integral in the complex plane 
\be I_N=\oint \frac{dw\, dz}{(2 \pi \rmi)^2} {1\over (z-w)^2} \prod_{k = 1}^N
\frac{(z + w - 2 z_k)^2}{( z- z_k)(w - z_k)}\ee
such that
\be I_N=2 D_{N-1} + 2^{2(N-1)}NI_1\ee
and
computing $I_N$ by expanding the contours of integration to infinity
\cite{paul}.

The eigenvalue grows slightly slower than linearly with $N$: 

\begin{center}
\begin{tabular}{|c|c|c|c|c|c|c|c|c|c|}
\hline
$N$ & 2 & 3 & 4 & 5 & 6 & 7 & 8 & 9 & 10 \\
\hline
$E_N$ & 1 & $\frac{9}{4}$ & $\frac{29}{8}$ & $\frac{325}{64}$
& $\frac{843}{128}$ & $\frac{4165}{512}$ & $\frac{9949}{1024}$
& $\frac{185517}{16384}$ & $\frac{424415}{32768}$ \\
\hline
\end{tabular}
\end{center}
The asymptotics is $E_N\to  2N - 4\sqrt\frac{N}{\pi}$.

Finally, as examples, here are a few other eigenstates.
For all $N$:
\be
\ft_N^{(1)} =
S_N \sum_{i=1}^N {\tilde z}_i^2\ee
with eigenvalue
\be\label{paulbis} E_N^{(1)} = E_N+\frac{1}{2^{2(N - 1)}}\frac{1}{\sum z_i^2}
 \oint \frac{dw\, dz}{(2 \pi \rmi)^2} \prod_{k = 1}^N
\frac{(z + w - 2 z_k)^2}{( z- z_k)(w - z_k)} =E_N+
\frac{1}{2^{2N - 2}} {2N \choose N - 1}
\ee
and 
\be
\ft_N^{(2)} =
S_N \sum_{i=1}^N {\tilde z}_i^3 \ee
 with eigenvalue
\bea
E_N^{(2)} & = &
 E_N+\frac{3}{ 2^{2(N - 1)}}\frac{1}{2\sum z_i^3}
 \oint \frac{dw\, dz}{(2 \pi \rmi)^2} (z + w) \prod_{k = 1}^N
\frac{(z + w - 2 z_k)^2}{( z- z_k)(w - z_k)} \nonumber \\
& = & E_N+
\frac{1}{2^{2N - 2}}\left[8 {2N -2 \choose N - 1} - {2 N + 1 \choose N - 1}\right] .
\label{paulter}
\eea
Both (\ref{paulbis}) and (\ref{paulter}) have been obtained by integration in the 
complex plane, following the procedure used in \cite{paul}.

For $N = 4$,  $L = 10$,   
one eigenstate is found to be a linear combination of
$h_1=\sum_{i<j}(z_i-z_j)^4$ 
and of the Pfaffian state
$h_2=(z_1-z_3)(z_1-z_4)(z_2-z_3)(z_2-z_4)-(z_1-z_2)(z_1-z_4)(z_2-z_3)(z_3-z_4)
+(z_1-z_2)(z_1-z_3)(z_2-z_4)(z_3-z_4)$.
Both $h_1$ and $h_2$ can be expanded in monomials of $\tilde z_i$ as
explained in the bosonic case.
Denoting $f_1=S_4 h_1$ and $f_2=S_4 h_2$ and introducing a matrix
$V_{ij}$ such that
\be \hat V f_i = \sum V_{ij}f_j \;,\ee 
one obtains 
\be
||V_{ij}|| = \left( \begin{array}{cc}
\frac{43}{16} & \frac{63}{32} \\
\frac{9}{64} & \frac{227}{128}
\end{array} \right) \;.
\ee
%The roots are solution of
%\be\label{tototer}   (227-128E)(86-32E)-63\times 18=0 \;:\ee
The eigenvalues are
\be
E_{4,10}^{(1,2)} = \frac{571 \pm 9\sqrt{393}}{256} \;.
\label{tototer}\ee

Clearly, the machinery developed above in the bosonic case can be
thoroughly used for  the fermionic model considered here. 
It follows that the $N$-body problem is
solvable in any given angular momentum sector. However, as in the
bosonic case, complicated irrational coefficients and eigenvalues are
expected, as illustrated in a particular case (\ref{tototer}).

\section{Conclusion}

Considering the quantum-mechanical model of bosons with a delta-function
coupling projected on the lowest Landau level, we have completely solved
the three-body problem, identified some analytic eigenstates for $N\ge4$
which belong to two hierarchies (the $L=5$ one is new),
and worked out an algorithm through which all other
eigenstates can be obtained by means of diagonalizing finite matrices
(i.e., they are solutions of algebraic equations of finite power).
An exact analytic solution of the $N$-body problem is evidently out of reach,
but a numerical solution to any precision is quite straightforward.
A model of fermions whose ground state is known to be the Laughlin
FQHE wave function, has been analyzed along the same lines.
We have shown that the Slater determinants, for any number of particles,
are eigenstates with rational eigenvalues related to Catalan numbers,
and identified a few excited states.
Here too, all the levels are solutions of algebraic equations.

\section{Acknowledgements}

We thank Thierry Jolicoeur for drawing our attention to the Smith-Wilkin
eigenstates and for numerous discussions and suggestions. We would also like
to thank Gora Shlyapnikov for enlighting comments and conversations.
S.~Mat.~acknowledges the hospitality of the Laboratoire de Physique
Th\'eorique et Mod\`eles Statistiques Orsay and support by
RFBR grant N 04-02-17087.


\begin{thebibliography}{99}
 \bibitem{Anderson95}
M.N.~Anderson, J.R.~Ensher, M.R.~Matthews, C.E.~Wieman, E.A.~Cornell,
Science 269, 198 (1995).
\bibitem{Bradley95}
C.C.~Bradley, C.A.~Sacket, J.J.~Tollet, R.G.~Hulet,
Phys. Rev. Lett. 75, 1687 (1995).
\bibitem{Davis95}
K.B.~Davis, M.-O.~Mewees, M.R.~Andrews, N.J.~van~Druten, D.S.~Durfee,
D.M.~Kurn, W.~Ketterle, Phys. Rev. Lett. 75, 3969 (1995).
\bibitem{Mattews99}
M.R.~Matthews, B.P.~Anderson, P.C.~Haljan, D.S.~Hall, C.E.~Wieman,
E.A.~Cornell, Phys. Rev. Lett. 83, 2498 (1999).
\bibitem{Madison00}
K.W.~Madison, F.~Chevy, W.~Wohlleben, J.~Dalibard,
Phys. Rev. Lett. 84, 806 (2000).
\bibitem{Madison01}
K.W.~Madison, F.~Chevy, V.~Bretin, J.~Dalibard,
Phys. Rev. Lett. 86, 4443 (2001).
\bibitem{Abo01}
J.R.~Abo-Shaeer, C.~Raman, J.M.~Vogels, W.~Ketterle,
Science 292, 476 (2001).
\bibitem{Engels02}
P.~Engels, I.~Coddington, P.C.~Haljan, E.A.~Cornell,
Phys. Rev. Lett. 89, 100403 (2002).
\bibitem{anyon}
J.M.~Leinaas, J.~Myrheim, Nuovo Cimento 37B, 1 (1977);
G.A.~Goldin, R.~Menikoff, D.H.~Sharp, J.~Math.~Phys. 21, 650 (1980),
J.~Math.~Phys. 22, 1664 (1981);
F.~Wilczek, Phys.~Rev.~Lett. 48, 1144 (1982),
Phys.~Rev.~Lett. 49, 957 (1982).
\bibitem{Nanyons}
S.~Mashkevich, Int.~J.~Mod.~Phys. A7, 7931 (1992);
G.~Dunne, A.~Lerda, S.~Sciuto, C.A.~Trugenberger, Nucl.~Phys. B370, 601 (1992);
A.~Karlhede, E.~Westerberg, Int.~J.~Mod.~Phys. B6, 1595 (1992).
\bibitem{LLLanyons} A.~Dasni\`eres de Veigy, S.~Ouvry, Phys. Rev. Lett. 72, 600 (1994).
\bibitem{Bertsch99} G.F.~Bertsch, T.~Papenbrock, Phys. Rev. Lett. 83, 5412 (1999).
\bibitem{Smith00} R.A.~Smith, N.K.~Wilkin, Phys. Rev. A 62, 061602 (2000).
\bibitem{Papenbrock01} T.~Papenbrock, G.F.~Bertsch, Phys. Rev. A 63, 023616 (2001).
\bibitem{Korslund06} M.N.~Korslund, S.~Viefers, e-print cond-mat/0602620.
\bibitem{Haldane} see for example F.~Duncan and M.~Haldane in ``The Quantum Hall Effect'', Editors R. E. Prange and S. M. Girvin, Springer-Verlag, 303 (1990). 
\bibitem{paul} P. Zinn-Justin, private communication.
\end{thebibliography}
\end{document}